\documentclass[a4paper, 10pt, conference]{ieeeconf}      

\IEEEoverridecommandlockouts                              

\usepackage{graphicx}

\title{\LARGE \bf
Spoken Dialogue System Based on Attribute Vector\\for Travel Agent Robot*
}

\author{Motoyuki Suzuki$^{1}$, Shintaro Sodeya$^{2}$ and Taichi Nakamura$^{3}$
\thanks{*A part of this work was supported by JSPS KAKENHI Grant Number JP22H01749}
\thanks{$^{1}$Motoyuki Suzuki is with Faculty of Information Science and Technology, Osaka Institute of Technology, Osaka, Japan
  {\tt\small moto@m.ieice.org}}%
\thanks{$^{2}$Shintaro Sodeya is with Graduate School of Information Science and Technology, Osaka Institute of Technology, Osaka, Japan}%
\thanks{$^{3}$Taichi Nakamura is with Faculty of Information Science and Technology, Osaka Institute of Technology, Osaka, Japan}%
}

\begin{document}

\maketitle
\thispagestyle{empty}
\pagestyle{empty}

\begin{abstract}
  In this study, we develop a dialogue system for a dialogue
  robot competition. In the system, the characteristics of sightseeing
  spots are expressed as ``attribute vectors'' in advance, and the user
  is questioned on the different attributes of the two candidate spots.
  Consequently, the system can make recommendations based on 
  user intentions.

  A dialogue experiment is conducted during a preliminary round of 
  competition. The overall satisfaction score obtained is 40.1 out of 63
  points,
  which is a reasonable result. Analysis of the relationship between
  the system behavior and satisfaction scores reveals that satisfaction
  increases when the system correctly understands the user intention and
  responds appropriately. However, a negative correlation is observed
  between the number of user utterances and the
  satisfaction score. This implies that inappropriate responses reduce the
  usefulness of the system as a consultation partner.
\end{abstract}

\section{INTRODUCTION}
In recent years, information processing technology for speech and natural
language has advanced. Consequently, several spoken dialogue systems have come
into practical use, most of which are the so-called
``question-and-answer'' systems, such as AI speakers and
Siri\cite{siri};
however, a practical system that can be a human advisor is still lacking.

To realize such a dialogue system, dialogue robot competitions have
been held since 2021\cite{DialogCompe1,DialogCompe2}.
In these competitions, a robot acts as a clerk at a travel agency and
consults with customers regarding their travel destinations while making
decisions.
Another feature of these competitions is that 
the naturalness of the dialogue, including non-verbal expressions
(such as gestures and facial expressions), is evaluated using an actual robot.

We developed a spoken dialogue system to participate in the
competition held in 2022\cite{DialogCompe2}.
In this competition, a user was required to consult with the robot
regarding two sightseeing spots selected by the user in advance, and decide
which spot is preferable for visit.
Because the development period was limited, the following policy was
established, and development was performed.
\begin{itemize}
\item The robot asks numerous questions. By asking the user numerous
  questions, it understands
  the user intention, and recommends sightseeing spots based on the
  results. We believe that user satisfaction improves if the
  robot fully understands the user intentions.
\item The robot does not disrupt spoken dialogue.
  Even when the speech is misrecognized or the expressions are unintelligible,
  it does not break up the dialogue; instead, it responds
  ``Uh-huh, I see.'' and allows the dialogue to continue until its end.
\item Non-verbal expressions are kept to a minimum, and the primary goal
  is to improve the completeness of the spoken dialogue system.
\end{itemize}

In this study, we describe the details of the spoken dialogue system we
developed, analyze the evaluations of users who interacted with
the system in the competition, and discuss key points to improve
user satisfaction.
\section{Recommendation method reflecting user intention}
\subsection{Overview}
The basic idea of this system is to fully understand the user
intention and recommend a sightseeing spot with features that match
the user intention. To achieve this, an ``attribute vector''
representing the characteristics of each spot was defined in advance.
Next, the user intentions were ascertained through dialogue and
expressed as a ``user vector.''
Finally, it was compared with the ``attribute vectors'' of
the candidate spots to make a recommendation.

The details of the algorithm are as follows.
\begin{enumerate}
\item The characteristics for each sightseeing spot are extract
  from sightseeing information\footnote{It consists of two
    files, SightBasic.csv and SightOption.csv.}
  provided by the convention organizer and create an ``attribute vector.''
\item The ``attribute vectors'' of the two candidate spots are compared,
  and the user intentions are questioned about the attributes with different
  values. The answers are analyzed to create a ``user vector'' that represents
  the user intentions.
\item The obtained ``user vector'' is compared with two ``attribute vectors''
  of sightseeing spots, and the recommended spot is communicated to
  the user with reasons based on the result.
\end{enumerate}
The algorithm can make recommendations by focusing on the
differences between the two candidate spots, and a
highly satisfactory dialogue can be achieved by correctly reflecting the
user intentions.

When making a recommendation in this manner, 
the characteristics of sightseeing spots must be extracted and
expressed in an ``attribute vector.'' 
Sightseeing spots can have a variety of characteristics with a great
deal of variation.
To collect such information, manually
collect information on each spot in advance and accumulating it in the
database would be effective. However, this competition had a rule that stated:
``the system should be able to work as soon as a list of sightseeing spots
is given,'' and collecting information manually is not allowed.

Although the competition allowed crawling various web pages
and collecting information automatically, we decided to generate
``attribute vectors'' only from the information provided by the
convention organizers.
\subsection{Definition of attribute vector}
Attributes used in this system are as follows.
\begin{itemize}
\item Type of sightseeing spot\\
  Four attributes: Art museum; Park; Museum; Observatory
\item Facilities\\
  Three attributes: Whether admission is charged or free; Whether parking is
  available; Whether it can be enjoyed even in the rain.
\item Type of recommended customers\\
  Five attributes: Children; Ladies; Babies; Alone; Customers with pets.
\item Recommended season\\
  Four attributes: Spring; Summer; Autumn; Winter.
\end{itemize}

A program was created to automatically extract the attribute values
from the information. The attribute values were three:
applicable (``yes''), not applicable (``no''), and no information
(``do not care''). For attributes related to the
``type of sightseeing spot,'' ``yes'' was given to only one of the
four types, and ``no'' was given to the rest. For the attributes
related to ``facilities,'' ``yes'' or ``no'' were selected for each.
For the attributes related to
``type of recommended customers'' and ``recommended season,'' 
positive information, such as
``recommended for children,'' was provided, whereas
``not recommended for children'' was not provided.
Therefore, ``yes'' or ``do not care'' was set.
\subsection{Acquisition method of user vector}\label{att}
The purpose of this dialogue was to determine which of the two
sightseeing spots the user would prefer to visit.
Therefore, we found the differences in the attributes of the two
spots and asked the user which attribute value they preferred in
these attributes.
The ``user vector'' was created by the answers to represent the
user intentions.

The details of the algorithm are as follows.
\begin{itemize}
\item Advance preparation
  \begin{enumerate}
  \item For all attributes used in ``attribute vector,''
    define a list of questions to be asked if
    there is a difference in the value of the attribute.
    For example, if the attribute is about the ``type of sightseeing spot,''
    define questions such as ``Do you like art museums?'' For an attribute
    related to the ``type of recommended customers,'' define questions
    such as ``Will you go with your children?''
  \item For each question, prepare a set of keywords and
    the update rules to determine which attribute values of ``user vector''
    should be updated if and when a keyword appears in the user's response.
    For a single question, assume various responses and construct a list of
    as many
    keywords as possible that may appear in the responses.
    For example, for the question ``Will you go with your children?,''
    if the answer is ``yes,'' the attribute ``recommended for children''
    is set to ``yes,'' the other attributes related to
    the ``type of recommended customers'' are set to ``no,''
    and all other attributes are set to ``do not care.''

    Furthermore, for each of the possible answers from the user,
    an appropriate response sentence is defined,
    such as ``You are with your child;
    therefore, we would like to recommend a place for your child.''
  \end{enumerate}
\item In the dialogue
  \begin{enumerate}
  \item Before starting the dialogue, all attribute values in the
    ``user vector'' are set to ``do not care.''
  \item Compare two ``attribute vectors'' of the selected sightseeing spots
    and extract all attributes with different attribute values.
    Ask questions for each extracted attribute.
  \item Extract keywords from the user's answer and update attribute
    values in the ``user vector'' based on the update rules.
    In this step;
    \begin{itemize}
    \item If the update rule is ``yes,'' update the attribute value
      to ``yes.''
    \item If the update rule is ``no,'' update the attribute value to
      ``no'' only if the value was ``do not care'' before the update.
      If the value was ``yes'' before updating, then ignore it.
    \item If the update rule is ``do not care,'' ignore it.
    \end{itemize}
  \end{enumerate}
\end{itemize}
The algorithm realizes that the user intentions can be acquired by
focusing on the
difference between two selected sightseeing spots to
make appropriate recommendations.
\subsection{Recommendation method}\label{recommend}
After the ``user vector'' is obtained, it is used to
recommend the sightseeing spot that matches the user intentions.
However, in this competition, the sightseeing spot that the travel
agency wants to recommend is defined separately from the user intention
and one of the objectives was to induce the user to choose this spot
through dialogue.

To achieve this objective, the following three attribute sets
were calculated and the
recommendation method was changed according to the number of elements.
In the following explanation, let $V_{r}$ be the ``attribute vector''
of the spot recommended by the travel agency,
$V_{n}$ the vector of the other spot, and $V_{u}$ the ``user vector.''
\begin{itemize}
\item $M(V_{r}, V_{u})$\\
  A set of attributes that have the attribute value ``yes'' in both
  $V_{r}$ and $V_{u}$. The more these attributes are, the more
  the user intention matches that of the spot recommended by the travel agency.
\item $U(V_{r}, V_{u})$\\
  A set of attributes whose value in $V_{r}$ is ``yes'' and those
  in $V_{u}$ is ``no.''
  The more these attributes are, the less the user intention
  matches that of the recommended spot.
\item $U(V_{n}, V_{u})$\\
  A set of attributes whose value in $V_{n}$ is ``yes'' and those
  in $V_{u}$ is ``no.''
  The more these attributes are, the less the user intention
  matches that of the other spot.
\end{itemize}

Depending on the number of attributes in these three sets (the number of
attributes in set A is denoted by $|A|$), the following recommendations
are made.
\begin{itemize}
  \item $|M(V_{r}, V_{u})| \ge |U(V_{r}, V_{u})|$\\
    In this case, because the number of attributes that match the user
    intention is larger than the number of attributes that do not match
    the user intention, a sightseeing spot recommended by the travel
    agency is recommended.

    Hence, the attributes included in $M(V_{r}, V_{u})$ are
    listed as reasons. Furthermore, the attributes included in
    $U(V_{n}, V_{u})$ are provided to explain that
    other spot does not match the user intention and guide not to select it.
\item $|M(V_{r}, V_{u})| < |U(V_{r}, V_{u})|$\\
  In this case, the recommended spot does not match the user
  intention. First, the attributes included in $U(V_{r}, V_{u})$
  are provided to describe that the recommended spot does not
  match the user intention, and then explain
  preferable points in the recommended spot and unpreferable points
  in the other spot based on $M(V_{r}, V_{u})$ and $U(V_{n}, V_{u})$.
  Finally, the robot recommends the recommended spot to the user.
\item $M(V_{r}, V_{u}) = U(V_{r}, V_{u}) = \phi$\\
  In this case, the user did not answer ``yes'' or ``no'' to the
  attributes of the recommended spot.
  Because the user intention is unknown, 
  the user is encouraged to select the recommended spot by explaining the
  characteristics of the spot, such as ``Generally recommended for children.''
\end{itemize}

The algorithm can guide the user to select the
sightseeing spot recommended by the travel agency while reflecting
the user intention.
\section{Dialogue Strategy}
During the development of the spoken dialogue system, importance was placed
on continuing
the dialogue to the end and ensuring it does terminate prematurely.
Understanding user utterance is achieved based on whether the
user utterance includes the pre-defined keywords.
However, utterances that do not include any of the
keywords may be input owing to misrecognition of utterances or
unexpected topics.
In such cases, the system does not ignore the utterance or listen to it again
but rather responds as if the system understands its content by
saying, ``I see.'' Therefore, the system ignores the user's
utterance,
which may cause inconsistencies in subsequent dialogues (such as asking
questions about the information contained in the ignored utterance);
however, we did not take any measures to address this.
In situations where the same response is provided regardless of the content
of the user's utterance, such as a response to the initial greeting,
a timeout is set such that the dialogue continues even if no input is received
for a certain period.

The entire dialogue flow is designed as follows.
\begin{enumerate}
\item Greetings and confirmation of selected spots\\
  The system briefly introduces itself and confirms that the two pre-selected
  sightseeing spots are correct.
\item Introducing the spots and asking about the reasons\\
  For each of the two spots, the introductory text included in the
  information provided by the convention organizer is read aloud,
  followed by a question about why this spot was chosen.

  If the assumed keywords exist in the user's response,
  the ``user vector'' value is updated based on the update rule, and
  the predefined response is returned.
\item General questions\\
  Because all spots have attributes related to the recommended
  customers, the question ``Will you travel alone?'' is asked.
  The attribute values related to the recommended customers in
  the ``user vector'' are updated based on the responses.
\item Questions about attributes that differ between the two spots\\
  As described in Section \ref{att}, for attributes with different values in
  the two spots, questions are asked to confirm the user intention,
  and the ``user vector'' is updated.
\item Recommendation of the spot\\
  The recommended sightseeing spot is explained with reasons based on
  the algorithm in Section \ref{recommend}.
\item Question and answer\\
  The system asks if the user has any questions regarding the two spots.
  If the user asks something, keywords are extracted, and the
  answer is provided. If there are no assumed keywords, the answer is
  ``Sorry, this information has not been provided.''
  Questions are accepted until the user indicates no questions.
\item Final greeting\\
  The dialogue is ended with a final greeting.
\end{enumerate}

Because the dialogue has a time limit,
the dialogue is terminated when the time exceeds five minutes, even if it is in
the middle of a dialogue.
\section{Analysis of dialogue}
Dialogue experiments and evaluations were conducted on the system during a
preliminary round of competition. The number of participants was 32
(20 males and 13 females). The ages of the participants ranged
from teenagers to fifties (of which, 14 were in their twenties,
eight in their forties, five in their thirties, and three each in
their teenagers and fifties).
\subsection{Recognition results}
The results of each user utterance were analyzed. A total of 483 user
utterances were observed in the entire dialogue, of which 174 (36.0\%)
were correctly
understood by the system and responded as expected.
\begin{table}[t]
  \begin{center}
    \caption{Cause of incorrect responses}\label{table1}
    \begin{tabular}{|l|rr|}
      \hline
      VAD failure&139&(28.8\%)\\\hline
      Misrecognition in ASR&42&(8.7\%)\\\hline
      Lack of keywords&127&(26.3\%)\\\hline
      Out of topics&20&(4.1\%)\\\hline
      Others&7&(1.4\%)\\\hline
    \end{tabular}
  \end{center}
\end{table}
For the remaining user utterances, the causes of incorrect
responses were classified. The results are listed in Table \ref{table1}.
Note that some of the utterances were counted more than once because
they were related to more than one cause.

The data presented in Table \ref{table1} reveals that 139 utterances (28.8\%)
were voice activity
detection (VAD) errors in the speech recognition system. In VADs,
even if the user uttered an utterance, it was not recognized as a speech
segment and speech recognition was not performed. In the
scene where the timeout was set (69 utterances), the dialogue continued as
if nothing had occurred. However, in other situations
(70 utterances), the system did not react to anything, and the dialogue
continued when the user spoke again. Furthermore, 42 utterances (8.7\%)
were incorrectly recognized as user intentions owing to
speech recognition errors. These two errors were caused by the speech
recognition system. The Google Speech Recognition
Engine\cite{Google-ASR} was used in this system; this was 
expected to improve recognition accuracy in noisy
environments.

The next most common cause was missing keyword registrations (26.3\%).
This system extracted keywords from the user's utterances, and used them to
determine the user intentions and generate response sentences.
The keywords were manually registered by assuming the user's utterances
in advance, but this assumption was insufficient. Consequently, the appropriate
dialogue could not be realized. This can be significantly
prevented by
registering additional keywords as required; however, registering all
keywords is not realistic. This is the principal limitation of 
keyword-based systems. Note that in some cases, the notations of the registered
keywords were different from those of the recognition results
(such as hiragana and kanji), and the correct response could not be obtained
although the keywords were registered.
\subsection{Relationship between robot's reaction and user's satisfaction}
In the experiment, all subjects were given a questionnaire regarding
their impressions of the system\cite{DialogCompe2}. The
relationship between the content of the dialogue and user satisfaction
obtained from the questionnaire was analyzed to determine the
factors that affected the level of satisfaction.

The questionnaire asked about the level of satisfaction with nine items on
a 7-point scale and the total score of these items was defined as
the ``overall satisfaction score.'' The average score
was 40.1 out of 63 points, which is not high; however,
the system was evaluated to some extent.

The correlation coefficients between the overall satisfaction
score and some possible items were calculated to determine the types
of responses of the
system that increased satisfaction. The results are presented in
Table \ref{satis}.
\begin{table}[t]
  \begin{center}
    \caption{Correlation coefficients between types of response and satisfaction score}
    \label{satis}
    \begin{tabular}{|l|r|}
      \hline
      Number of user's utterances&$-0.37$\\\hline
      Percentage of appropriate response&$+0.42$\\\hline
      Percentage of incorrect response&$-0.02$\\\hline
      Percentage of the response ``I see''&$-0.19$\\\hline
      Number of restatements&$-0.29$\\\hline
    \end{tabular}
  \end{center}
\end{table}
      
Table \ref{satis} revealed that the percentage of appropriate responses
had a correlation coefficient of $+0.42$, thus
indicating a significant impact on the impression of the dialogue system.
Particularly, this percentage was correlated with several items
in the questionnaire:
 ``Were you satisfied with the choice of tourist attractions
to visit?'' ($+0.48$); ``Did you trust the robot?'' ($+0.41$), and
``Did you use the information obtained from the robot to select the
sightseeing spot?'' ($+0.41$). These results also show that the dialogue
system was trusted and recognized as worth consulting.

However, the number of utterances by the user 
had a negative correlation coefficient of $-0.37$. As the number of
utterances here included ``restatements'' (the user repeats the same
utterance because the system did not react to it despite the user's
utterance), restatements were assumed to reduce satisfaction.
However, the correlation coefficient between the number of restatements
and overall satisfaction score was $-0.29$; therefore, 
this could not be concluded as the only cause.
As the number of user utterances increases,
the number of inappropriate responses by the system may also increase,
thereby reducing the level of satisfaction. 
Note that the number of restatements also had a higher negative correlation
with questions ``Were you able to obtain sufficient
information about the sightseeing spots?'' ($-0.40$) and
``Would you like to visit this travel agency again?'' ($-0.38$),
thus indicating a negative impact on the objective of consulting.

Moreover, neither the percentage of incorrect responses nor ``I see''
responses exhibited a significant correlation with the overall satisfaction
score. Although a more detailed analysis is required, 
it does not seem to directly lower the level of satisfaction.
\section{CONCLUSIONS}
In this study, we proposed a spoken dialogue system that acts as a travel
agent and allows the user to consult with the agent regarding the selection
of sightseeing spots. By examining the differences in characteristics
between the candidate spots and asking the user about their intention,
the system can recommend a sightseeing spot that is more in line with
the user intention. Additionally, even when the system does not understand
what is being said, it does not ask the user to repeat what they
said; instead it responds as if it understood,
thus avoiding a break in the dialogue.

A dialogue experiment was conducted in the preliminary round of a
spoken dialogue competition, and an overall satisfaction score of
40.1 out of 63 points, which is a reasonable score, was obtained.
Evidently, satisfaction increased significantly when the system was able to
respond
appropriately to the user's speech; thus indicate responding in a manner
whereby the user feels that the system has correctly
understood the speech is crucial.
\addtolength{\textheight}{-20cm}   



%
%


\bibliographystyle{IEEEtran}
\bibliography{/home/moto/tex/macros/bibdata}
\end{document}